\documentclass[aps,twocolumn]{revtex4}
\usepackage{epsfig}
\usepackage{psfrag}
\newcommand{\be}{\begin{equation}}
\newcommand{\ee}{\end{equation}}
\newcommand{\bea}{\begin{eqnarray}}
\newcommand{\eea}{\end{eqnarray}}
\newcommand{\ba}{\begin{array}}
\newcommand{\ea}{\end{array}}
\newcommand{\E}{{\sf E}}

\begin{document}

\title{Entropy Estimates from Insufficient Samplings}

\author{Peter Grassberger}
\affiliation{Complex Systems Research Group, John-von-Neumann Institute for Computing,
J\"ulich Research Center, D-52425 J\"ulich, Germany}

\date{\today}

\begin{abstract}
We present a detailed derivation of some estimators of Shannon entropy for 
discrete distributions. They hold for finite samples of $N$ points distributed
into $M$ ``boxes", with $N$ and $M \to\infty$, but $N/M < \infty$. In the high 
sampling regime ($\gg 1$ points in each box) they have exponentially small 
biases. In the low sampling regime the errors increase but are still much
smaller than for most other estimators. One advantage is that our main 
estimators are given analytically, with explicitly known analytical formulas for
the biases.
\end{abstract}

\maketitle

It is well known that estimating (Shannon) entropies from finite samples
is not trivial. If one naively replaces the probability $p_i$ to be in 
``box" $i$ by the observed frequency, $p_i \approx n_i/N$, statistical
fluctuations tend to make the distribution look less uniform, which 
leads to an underestimation of the entropy. There have been numerous
proposals on how to estimate the bias 
\cite{miller,harris,herzel,grass88,schmitt,wolpert,poschel,panzeri,schuermann,strong,holste,nemenman,paninski}. 
Some make quite strong assumptions \cite{schmitt,poschel}, others use Bayesian
methods \cite{wolpert,holste,nemenman}. 
As pointed out in \cite{grass88,paninski}, one can devise estimators with 
arbitrarily small bias (for sufficiently large $N$ and fixed $p_i$), but these 
will then have very large statistical errors (if sufficiently many of the $n_i$ 
are small but $\neq 0$). In the present paper we want to 
revisit a method used in \cite{grass88}. There a very simple correction term 
was derived which seems to be a very good compromise between bias, statistical
errors, and ease of use. Unfortunately, the treatment in \cite{grass88} was 
not quite systematic, and in particular the corrections going beyond the 
proposed term were wrong. It is the purpose of the present letter to 
provide a more systematic presentation of the method used in \cite{grass88},
to correct some of the errors made there, and to propose an estimator which
is again very easy to use and which should be better than that proposed 
in \cite{grass88}.

We consider $M \gg 1$ ``boxes" (states, possible experimental outcomes, ...) 
and $N \gg 1$ points or particles distributed randomly and independently 
into the boxes. We assume that each box has weight $p_i$ ($i=1,\ldots M$)
with $\sum_i p_i = 1$. Thus each box $i$ will contain a random number $n_i$
of points, with $\E[n_i] = p_iN$. Their distribution is binomial, 
\be
   P(n_i;p_i,N) = {N\choose n_i} p_i^{n_i} (1-p_i)^{N-n_i} .    \label{binomi}
\ee
Since entropy $H$ is a sum over terms each of which depends only on 
one index $i$, we only need these marginal distributions instead of 
the more complicated and non-factorizing joint distribution.
Some of the $p_i$ can be zero, but in the following we shall assume that
none of them is large, i.e. $p_i \ll 1$ for all $i$. In that limit 
the numbers $n_i$ are Poisson distributed,
\be
   P(n_i;z_i) = {z_i^{n_i}\over n_i!} e^{-z_i}         \label{poiss}
\ee
with $z_i \equiv \E[n_i] = p_iN$. The error in going from Eq.(\ref{binomi})
to (\ref{poiss}) is $O(1/N)$. Thus all derivations given below hold strictly 
only in the limit $N\to\infty,\; M\to\infty,\; n_i/N\to 0 \;\;\forall i$, but 
the general case is not much more difficult, see footnote \cite{foot}.

Our aim is to estimate the entropy, 
\be
   H = -\sum_{i=1}^M p_i \ln p_i = \ln N - {1\over N}\sum_{i=1}^M z_i \ln z_i,
\ee
from an observation of the numbers $\{n_i\}$ (in the following, all 
entropies are measured in ``natural units", not in bits). The estimator 
$\hat{H}(n_1,\ldots n_M)$ will of course have both statistical 
errors and a bias, i.e. if we repeat this experiment, the average
of $\hat{H}$ will in general not be equal to $H$,
\be
   \Delta H \equiv \E[\hat{H}] - H \neq 0.
\ee
In the limit $N\to\infty, M\to\infty$, the statistical error will
go to zero (because essentially one averages over many boxes), but 
the bias will remain finite unless also $n_i\to\infty\;\;\forall i$ in this 
limit, which we will not assume in the following.
Indeed it is well known that the {\it naive estimator}, obtained by 
assuming $z_i = n_i$ without fluctuations,
\be
   \hat{H}_{naive} = \ln N - {1\over N}\sum_{i=1}^M n_i \ln n_i,
\ee
is negatively biased, $\Delta H_{naive} < 0$.

In the limit of large $N$ and $M$ each contribution $z_i \ln z_i$ to the 
entropy will be statistically independent, and can thus also be 
estimated independently by some estimator which is only a function of 
$n_i$ \cite{grass88}, 
\be
   z_i \ln z_i \approx \widehat{z_i \ln z_i} = n_i\phi(n_i)
   \label{zi}
\ee
such that its expectation value is
\be
   \E[\widehat{z_i \ln z_i}] = \sum_{n_i = 1}^\infty n_i\phi(n_i) P(n_i;z_i).
   \label{Ezi}
\ee
Notice that the sum here runs only over strictly positive values
of $n_i$. Effectively this means that we have assumed that observing 
an outcome $n_i=0$ does not give any information: If $n_i=0$, we do not 
know whether this is because of statistical fluctuations or because $p_i=0$ 
for that particular $i$.

The resulting entropy estimator is then \cite{grass88} \cite{foot}
\be
   \hat{H}_\phi = \ln N - {M\over N} \overline{n\phi(n)}  \label{8}
\ee
with the overbar indicating an average over all boxes,
\be
   \overline{n\phi(n)} = {1\over M} \sum_{i=1}^M n_i \phi(n_i).
\ee
Its bias is
\be
   \Delta H_\phi = {M\over N} (\overline{z\ln z} -\E[\overline{n\phi(n)}]).
   \label{dHphi}
\ee

It will turn out that some of the derivations given below simplify if
we consider instead of the Shannon case the more general Renyi 
entropies,
\bea
   H(q) & = & {1\over 1-q} \ln \sum_{i=1}^M p_i^q \nonumber \\ 
        & = & {1\over 1-q} [ \ln \sum_{i=1}^M z_i^q - q \ln N].
   \label{Renyi}
\eea
The Shannon case is recovered by taking the limit $q\to 1$, 
$H = \lim_{q\to 1}H(q)$. Eqs.(\ref{zi}) to (\ref{dHphi}) are then
replaced by $ \widehat{z_i^q} = n_i\phi(n_i,q)$ with 
$\phi(n) = d\phi(n,q)/dq|_{q=1}$, $\E[\widehat{z_i^q}] = \sum_{n_i} 
n_i\phi(n_i,q) P(n_i;z_i)$, and 
\be
   \Delta \exp((1-q)H(q))_\phi = 
{M\over N} (\overline{z^q} - \E[\overline{n\phi(n,q)}]).
\ee

For integer $q \geq 2$, the bias-free estimator is given by 
(in the following we shall suppress the index $i$)
\be
   \widehat{z^q} = {n!\over (n-q)!},
\ee
since the {\it factorial moments} satisfy \cite{foot}
\be
   \sum_{n=q}^\infty {n!\over (n-q)!} P(n;z) = z^q. \label{factorial}
\ee
This suggests that it might be a good strategy to look first at the 
generalization of the l.h.s. for arbitrary $q$, and then analyze
more closely the difference with $z^q$. In addition, we will see
that we should start with {\it negative} real $q$, and go to 
positive $q$ only later by analytic continuation.

We thus define 
\bea
   A(-q,z) & = & \sum_{n=1}^\infty {\Gamma(n+1)\over\Gamma(n+1+q)} {z^n\over n!} e^{-z}
   \nonumber \\
      & \equiv & \E[{\Gamma(n+1)\over \Gamma(n+1+q)}].
\eea
We write $\Gamma(n+1)/\Gamma(n+1+q) = B(n+1,q)/\Gamma(q)$ and use 
the integral representation for the beta function (Ref.\cite{abramow}, 
paragraph 6.2.1)
\be
   B(n+1,q) = \int_0^1 \;dt\; (1-t)^n t^{q-1} .
\ee
Since both this integral and the sum over $n$ in the definition
of $A(-q,z)$ are absolutely convergent, we can interchange them.
The sum can then be done exactly, giving
\bea
   A(-q,z) & =& {1\over \Gamma(q)} \int_0^1\; dt\; t^{q-1} (e^{-tz} - e^{-z})  \nonumber \\
    & = & {z^{-q}\over \Gamma(q)} \int_0^z \;dx\; x^{q-1}e^{-x} - {e^{-z}\over \Gamma(1+q)}.
\eea
The last term arises since the sum over $n$ extends only from 1 to $\infty$.
Writing now $\int_0^z = \int_0^\infty - \int_z^\infty$ we can express the 
first term as a Gamma function and the second as an incomplete Gamma 
function (\cite{abramow}, paragraph 6.5.3),
\be
   A(-q,z) = z^{-q} - {z^{-q}\over \Gamma(q)} \Gamma(q,z) - {e^{-z}\over \Gamma(1+q)}.
\ee
Here we can finally continue analytically to positive $q$. Furthermore
we use the recursion relation (Ref.\cite{abramow}, paragraph 6.5.22)
\be
   \Gamma(a,z) = {1\over a} \Gamma(1+a,z) - {e^{-z} z^a\over a}
\ee
to arrive finally at 
\be
  \E[{\Gamma(n+1)\over \Gamma(n+1-q)}] = z^q - {z^q\over \Gamma(1-q)} \Gamma(1-q,z).
\ee
For the Shannon case we take the derivative with respect to $q$ at $q=1$ and
obtain \cite{foot}
\be
  \E[n\psi(n)] = z\ln z + z E_1(z)\;. \label{psi}
\ee
Here, $\psi(x) = d\ln \Gamma(x)/dx$ is the digamma function, and 
\be
   E_1(x)=\Gamma(0,x) = \int_1^\infty {e^{-xt}\over t} dt
\ee
is an exponential integral (Ref.\cite{abramow}, paragraph 5.1.4).

Eq.(\ref{psi}) is our first important result. For large values of $z$,
$z E_1(z)\approx e^{-z}$. Thus, if $z = \E[n]$ is large, it is an exponentially
good approximation to simply neglect the last term in Eq.(\ref{psi}). 
We call the resulting entropy estimator $\hat{H}_\psi$ \cite{grass88} \cite{foot},
\be
   \hat{H}_\psi = \ln N - {1\over N}\sum_{i=1}^M n_i\psi(n_i). \label{hpsi}
\ee
Moreover, for $z\to 0$ we have also $z E_1(z)\to 0$, and in between 0 and
$\infty$ the function is positive with a single maximum at $z=0.434...$
where $zE_1(z) = 0.2815...$. If we simply neglect the last term, we make
thus a negative bias, but at most by 
\be
   0 < - \Delta H_\psi = \overline{z E_1(z)} M/N < 0.2815\ldots\;\times\; M/N.
\ee
If we approximate further $\psi(x)\approx \ln x$, we obtain the naive
estimator. The better approximation $\psi(x)\approx \ln x -1/2x$ gives
Miller's correction \cite{miller,herzel}. It can be shown that 
\be
   \E[n\ln n] > \E[n\ln n -1/2] > \E[n\psi(n)] > z\ln z
\ee
for all positive $z$.
Thus both the naive estimate and Miller's correction are worse than $\hat{H}_\psi$.
The difference is especially big for large $z$, where the error of the naive 
estimate goes to $M/2N$, the error after applying Miller's correction is $\sim
M/zN$, while the error of $\hat{H}_\psi$ is $\sim \exp(-z)M/N$.

But we can do even better. First we notice that 
\be
   -\sum_{n=1}^\infty{(-1)^n\over n+1}{z^n\over n!} e^{-z} = 
      e^{-z}-{e^{-z}\over z} + {e^{-2z}\over z}
\ee
which has the same leading behaviour for large $z$ as $zE_1(z)$. 
It also goes to zero for $z\to 0$, is positive for all $z\in [0,\infty)$,
and is smaller than $zE_1(z)$ for all $z$. Thus, replacing 
$\psi(n)$ by 
\be 
   \psi(n)+{(-1)^n\over n(n+1)}     \label{oneovern}
\ee
gives an improved estimator.
Apart from a misprint, this is the estimator recommended in 
\cite{grass88}, Eq.(13).

This equation had been derived in \cite{grass88} somewhat unsystematically, 
using asymptotic series expansions in an uncontrolled way. Because of 
that, the discussion of the more general approximation, Eq.(11)
in that paper, is wrong. In particular, Eq.(11) holds (for $q\to 1$)
not for all integer $R$, but only for odd values of $R$. Furthermore,
the fact that the terms neglected in Eq.(11) decrease as $z^{-R}e^{-z}$ for 
large $z$ does {\it not} mean that Eq.(11) is exact in the limit 
$R\to\infty$. Finally, in contrast to what is said there, this limit
can be taken without a risk of statistical errors blowing up, at least 
for $q\to 1$. 

Instead of following the derivation of \cite{grass88}, we consider
the semi-infinite sequence of real numbers $G_1, G_2, \ldots$ 
defined by 
\bea
    G_1 & = & -\gamma - \ln 2,  \nonumber \\
    G_2 & = & 2-\gamma - \ln 2,  \nonumber \\
    G_{2n+1} & = & G_{2n},
\eea
(here, $\gamma = 0.577215\ldots$ is Euler's constant) and 
\be
    G_{2n+2} = G_{2n}+{2\over 2n+1} \quad (n\geq 1).
\ee
Thus $G_{2n} = -\gamma - \ln 2 + 2/1+2/3+2/5+\ldots + 2/(2n-1)$. Using 
the representation $\psi(n) = -\gamma + 1/1+1/2+1/3+\ldots + 1/(n-1)$,
one checks that 
\be
   G_n = \psi(n) + (-1)^n\int_0^1{x^{n-1}\over x+1} dx.
\ee
On the one hand, using formula 0.244 of \cite{gradshteyn}, one can 
write this integral as an infinite sum,
\be
   G_n = \psi(n) + (-1)^n\sum_{l=0}^\infty {1\over (n+2l)(n+2l+1)},
  \label{sumG}
\ee
which can be compared to Eq.(11) of \cite{grass88} with $q\to 1$
and odd $R\to\infty$. On the other hand, we obtain
\bea
   \E[n(G_n-\psi(n))]
       & = & \sum_{n=1}^\infty n(G_n-\psi(n)) {z^n\over n!} e^{-z} \nonumber\\ 
       & = & -\int_0^1{dx\over x+1} e^{-z} z \sum_{n=1}^\infty {(-xz)^{n-1}
              \over (n-1)!}  \nonumber\\ 
       & = & - e^{-z} z \int_0^1{dx\over x+1} e^{-xz} \nonumber\\ 
       & = & -z(E_1(z)-E_1(2z)).
\eea
Therefore, combining this with Eq.(\ref{psi}), we have 
\be
   \E[nG_n] = z\ln z + z E_1(2z).
\ee
This is our main result. Since the last term decreases as $e^{-2z}$, 
the error made when neglecting it decreases exponentially faster 
with $z=\E[n]$ than when neglecting the last term in Eq.(\ref{psi}),
for large $z$. Thus, if all boxes have $\E[n_i]>5$, say, the error
committed is $<e^{-10}$ which should be negligible in all practical
cases. More generally, the error made by neglecting the last term
is again always negative, and it is bounded by 
\be
   0 < - \Delta H_G < 0.1407\ldots\;\times \; M/N,        \label{error2}
\ee
where \cite{foot}
\be
   \hat{H}_G = \ln N - {1\over N} \sum_{i=1}^M n_i G_{n_i}
    \label{twoz}
\ee
is our proposed best estimator. 

\begin{figure}
  \begin{center}
  \psfig{file=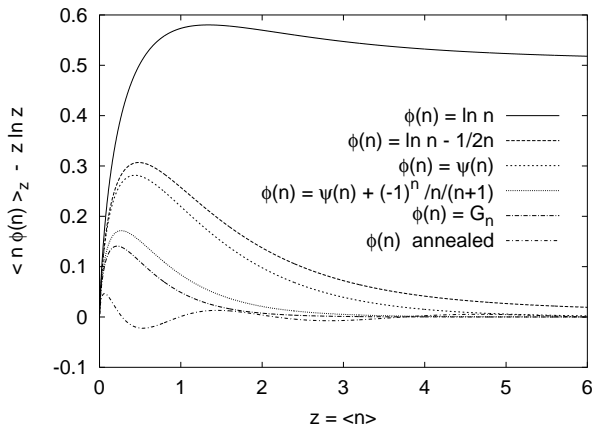,width=5.8cm, angle=270}
   \caption{Error terms for fixed $z=\E[n]$ and for different 
     functions $\phi(n)$. While the first five are analytic, the
     last one is just one typical simulated annealing result. 
     Different cost functions, annealing schemes, and random 
     number sequences will give slightly different results.}
    \label{error}
  \end{center}
\end{figure}

\begin{figure}
  \begin{center}
  \psfig{file=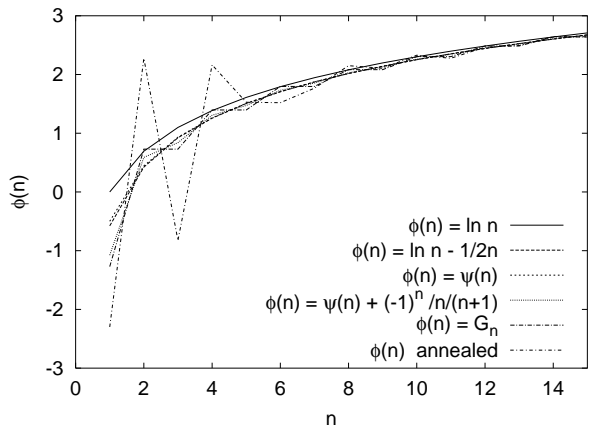,width=5.8cm, angle=270}
   \caption{Functions $\phi(n)$ corresponding to the error terms 
     shown in Fig.1. Notice that they are defined only for 
     integer $n$. Values at non-integer $n$ are just plotted 
     to guide the eye.}
    \label{phi}
  \end{center}
\end{figure}

Let us denote by $z^*=0.217\ldots$ the position of the maximum of 
$zE_1(2z)$. For $z<z^*$ this function is convex. Thus, if 
$N/M < z^*$, the distribution of $z$-values over the boxes which 
gives the maximal bias is a delta function, $P(z) = \delta(z-N/M)$, 
and Eq.(\ref{error2}) can be improved to $-\Delta H_G \leq 
E_1(2N/M)$. For $N/M\to 0$ this diverges $\sim \ln (M/N)$.

We might add that truncating the sum in Eq.(\ref{sumG}) at any 
finite $l$ also gives valid estimators whose errors are between 
those of $\hat{H}_G$ and $\hat{H}_\psi$, but there seems no reason 
to prefer any of them over $\hat{H}_G$ or $\hat{H}_\psi$. Taking 
only the term with $l=0$ gives Eq.(\ref{oneovern}).

The error terms $\E[n\phi(n)] - z\ln z$ for $\phi(n) = \ln n,\; \ln n -1/2,\;
\psi(n),\;\psi(n)+(-1)^n/n/(n+1)$, and $G_n$ are shown in Fig.1, together 
with one more curve discussed below. The functions $\phi(n)$ themselves
are shown in Fig.2.

We can give estimators with even smaller absolute bias, i.e. with 
$|\Delta H| < 0.1407\ldots$, but they have several drawbacks:

\begin{itemize}
\item Their biases can have either sign. 
\item We were only able to find them numerically, by minimizing
(by simulated annealing) a cost function like e.g. the $L^2$ norm
\be
   \delta = \int_0^\infty \;{dz \over\sqrt{z}}\;
      | \sum_{n=1}^\infty n\phi(n) {z^n\over n!} e^{-z} -z\ln z|^2.
\ee
Typical results obtained in this way are shown in Figs.1 and 2 
\cite{foot2}.
\item The resulting function $\phi(n)$ replacing $\psi(n)$ resp.
$G_n$ is not monotonic, and its total variation as measured e.g. by
$\sum_{n=1}^\infty \;n\;[\phi(n)-G_n]^2$ would diverge as $\delta\to 0$
(indeed, the results shown in Figs.1 and 2 were obtained by adding 0.0002 
times this term as a regularizer to the $L^2$ norm). This is 
the most serious drawback. It means that large cancellations must occur 
and thus {\it statistical} errors blow up in the limit $\delta\to 0$ 
(if $N$ is kept finite), 
as is to be expected on general grounds \cite{grass88}. There cannot be 
any estimator of $H$ completely free of bias for finite $N$. Notice that 
$G_n$ is the ``best" sequence which is still monotonic. Estimates based on 
non-monotonic $\phi(n)$ might be useful if one has important contributions 
from extremely small $z$, i.e. if either $N\ll M$ or if the distribution 
of $p_i$ is so uneven that many boxes have small (but not too small) $z_i$. 

\end{itemize}

I have applied the above estimators to the six examples shown in Fig.4 of \cite{nemenman}.
In each of these examples the number of boxes was $M\geq 1000$, although
the number of non-empty boxes was smaller in some of them. Nevertheless,
the distributions were severely undersampled in most cases when 
$N\leq 300$. In all cases the annealed $\phi(n)$ shown in Fig.2 gave 
statistical errors smaller or comparable to the 
Bayesian estimators of \cite{nemenman}, and the bias was smaller than 
the statistical errors for all $N\geq 300$. In all but two cases
(Zipf's law and $\beta=1$, with $\beta$ defined in \cite{nemenman}) 
the bias was negligible even down to $N=10$.
With Eq.(\ref{twoz}), the bias was significant ($>2\sigma$) in the same 
two cases for all $N\leq 300$, and in the case $\beta=0.02$ for $N=10$.
   
In summary, I hope to have clarified the arguments and corrected 
the mistakes made in \cite{grass88}, and I have substantially improved
on the results. I have proposed a new analytic
estimator for Shannon entropy which has very small systematic errors,
except when the average number of points per box is much smaller than
1. Its statistical errors should be larger than those of the naive 
estimator (since there contributions from $n_i=1$ and from $n_i>1$
partially cancel), but this difference should be small. In addition,
it is shown that numerically obtained estimators can be useful for 
extremely undersampled cases.
The estimator $\hat{H}_\psi$ and the first correction based on 
Eq.(\ref{oneovern}) can be generalized straightforwardly to Renyi
entropies, but I was not able to generalize the new estimator,
Eq.(\ref{twoz}), to $q\neq 1$. The present estimators can not match the 
best Bayesian estimators \cite{nemenman} when the sampling is extremely
low, but they are much simpler to use and 
more robust, as no guess of any prior distribution is needed. 

I want to thank Walter Nadler for carefully reading the manuscript, and to 
Liam Paninski for correspondence.


\begin{thebibliography}{99}
\bibitem{miller} G. Miller, Note on the bias of information estimates.
   In H. Quastler, ed., {\it Information theory in psychology II-B}, 
   pp 95-100 (Free Press, Glencoe, IL 1955).
\bibitem{harris} B. Harris, Colloquia Math. Soc. Janos Bolya, p. 323 (175).
\bibitem{herzel} H. Herzel, Sys. Anal. Mod. Sim. {\bf 5}, 435 (1988).
\bibitem{grass88} P. Grassberger, Phys. Lett. {\bf A 128}, 369 (1988).
\bibitem{schmitt} A.O. Schmitt, H. Herzel, and W. Ebeling, Europhys. Lett.
   {\bf 23}, 303 (1993).
\bibitem{wolpert} D.H. Wolpert and D.R. Wolf, Phys. Rev. E {\bf 52}, 6841 (1995). 
\bibitem{poschel} T. Poschel, W. Ebeling, and H. Rose, J. Stat. Phys.
   {\bf 80}, 1443 (1995).
\bibitem{panzeri} S. Panzeri and A. Treves, Network: Computation in
     Neural Systems  {\bf 7}, 87 (1996). 
\bibitem{schuermann} T. Sch\"urmann and P. Grassberger, Chaos {\bf 6}, 414 (1996).
\bibitem{strong} S. Strong, R. Koberle, Rob R. de Ruyter van Steveninck, and W. Bialek,
   Phys. Rev. Lett. {\bf 80}, 197-200 (1998).
\bibitem{holste} D. Holste, I. Grosse, and H. Herzel, J. Phys. A {\bf 31},
   2551 (1998). 
\bibitem{nemenman} I. Nemenman, F. Shafee, and W. Bialek, Entropy and inference,
   revisited. In T.G. Dietterich {\it et al.}, eds., {\it Advances in neural
   information processing 14} (MIT Press, Cambridge 2002).
\bibitem{paninski} L. Paninski, Neural Computation {\bf 15}, 1191 (2003)
\bibitem{foot} For the correct binomial distribution, Eq.(\ref{factorial}) is replaced 
   by $\E[n!/(n-q)!] = p^q N!/(N-q)!$, and Eq.(\ref{psi}) by 
   \begin{displaymath}
     \qquad \E[n\psi(n)]  =  z\ln z + z[\psi(N)-\ln N]  
                   +  z\int_0^{1-p}{x^{N-1}dx\over 1-x} \;.
   \end{displaymath}
   The last term can be estimated very similarly to the last term in 
   Eq.(\ref{psi}), in particular it is positive and is bounded for all $N$ and $p$ 
   by ${N\over N-1}(1-p)^N$. In the estimator Eq.(\ref{hpsi}) which results 
   from neglecting this term, replacing the Poisson distribution by the correct 
   binomial one amounts to replacing $\ln N$ by $\psi(N)$ (bringing e.g. 
   Miller's correction from $M/2N$ down to $(M-1)/2N$ \cite{paninski}). 
   Similarly, in Eq.(\ref{twoz}) one should replace $\ln N$ by $G_N$, in 
   order to correct for the most important $O(1/N)$ term not included in the
   Poisson approximation. For all estimators (including those where $\phi(n)$
   is obtained numerically), one should replace in Eq.(\ref{8}) $\ln N$ by
   $\phi(N)$.
\bibitem{abramow} M. Abramowitz and I. Stegun, eds., {\it 
   Handbook of Mathematical Functions} (Dover, New York 1965).
\bibitem{gradshteyn} I.S. Gradshteyn and I.M. Ryshik, {\it Tables of 
   Integrals, Series, and Products} (Academic Press, New Yrok 1965).
\bibitem{foot2} The coefficients $\phi(n)$ for this solution can be obtained by 
   sending an e-mail to \verb/p.grassberger@fz-juelich.de/.
\end{thebibliography}
\end{document}